\def\year{2016}
\documentclass[letterpaper]{article}
\usepackage{aaai16}
\usepackage{times}

\usepackage{url}
\usepackage{amsmath}
\usepackage{amssymb}
\usepackage{graphicx}

\usepackage{amssymb}
\usepackage{algorithm}
\usepackage{algorithmicx}
\usepackage{algpseudocode}
\usepackage{graphicx}
\usepackage{amsmath,bm}
\usepackage{subfigure} 
\usepackage{amsmath}
\usepackage{wrapfig}
\usepackage[]{xspace}

\newcommand{\bxi}{\ensuremath{\boldsymbol{\xi}}}
\newcommand{\cov}{\ensuremath{\Sigma_u}}
\newcommand{\bx}{\ensuremath{{\mathbf{x}}}}

\newcommand{\bv}{\ensuremath{{\mathbf{v}}}}
\newcommand{\bu}{\ensuremath{{\mathbf{u}}}}
\newcommand{\bof}{\ensuremath{{\mathbf{f}}}}
\newcommand{\bG}{\ensuremath{{\mathbf{G}}}}
\newcommand{\bR}{\ensuremath{\mathbf{R}}}
\newcommand{\btau}{\ensuremath{\boldsymbol{\tau}}}
\newcommand{\citet}[1]{\citeauthor{#1}~{\shortcite{#1}}}
\newcommand{\citett}[2]{\citeauthor{#1}~\shortcite{#1}}

\newcommand{\qrsim}{\texttt{QRSim}\xspace}
\newcommand{\sname}{\texttt{CRATES}\xspace}

\usepackage{lineno}

\begin{document}

\title{Real-Time Stochastic Optimal Control for Multi-agent Quadrotor Systems}
\author{Vicen\c{c} G\'omez$^1$, Sep Thijssen$^2$, Andrew Symington$^3$, Stephen Hailes$^4$, Hilbert J. Kappen$^2$
\vspace{.25cm}
\\
$^1$Universitat Pompeu Fabra. Barcelona, Spain~~\texttt{vicen.gomez@upf.edu}\\
$^2$Radboud University Nijmegen, the Netherlands~~\texttt{\textbraceleft s.thijssen,b.kappen\textbraceright @donders.ru.nl}\\
$^3$University of California Los Angeles, USA~~\texttt{andrew.c.symington@gmail.com}\\
$^4$University College London, United Kingdom~~\texttt{s.hailes@cs.ucl.ac.uk}
}

\maketitle
\begin{abstract}
This paper presents a novel method for controlling teams of unmanned aerial vehicles using Stochastic Optimal Control (SOC) theory.
The approach consists of a centralized high-level planner that computes optimal state trajectories as velocity sequences, and a platform-specific low-level controller which ensures that these velocity sequences are met.
The planning task is expressed as a centralized path-integral control problem, for which optimal control computation corresponds to a probabilistic inference problem that can be solved by efficient sampling methods. Through simulation we show that our SOC approach (a) has significant benefits compared to deterministic control and other SOC methods in multimodal problems with noise-dependent optimal solutions, (b) is capable of controlling a large number of platforms in real-time, and (c) yields collective emergent behaviour in the form of flight formations. Finally, we show that our approach works for real platforms, by controlling a team of three quadrotors in
outdoor conditions.
\end{abstract}

\section{Introduction}
The recent surge in autonomous Unmanned Aerial Vehicle (UAV) research
has been driven by the ease with which platforms can now be acquired,
evolving legislation that regulates their use, and the broad range
of applications enabled by both individual platforms and cooperative
swarms. Example applications include automated delivery systems, monitoring
and surveillance, target tracking, disaster management and navigation
in areas inaccessible to humans.

\begin{figure}[t!]
\centering{}\includegraphics[width=.95\columnwidth]{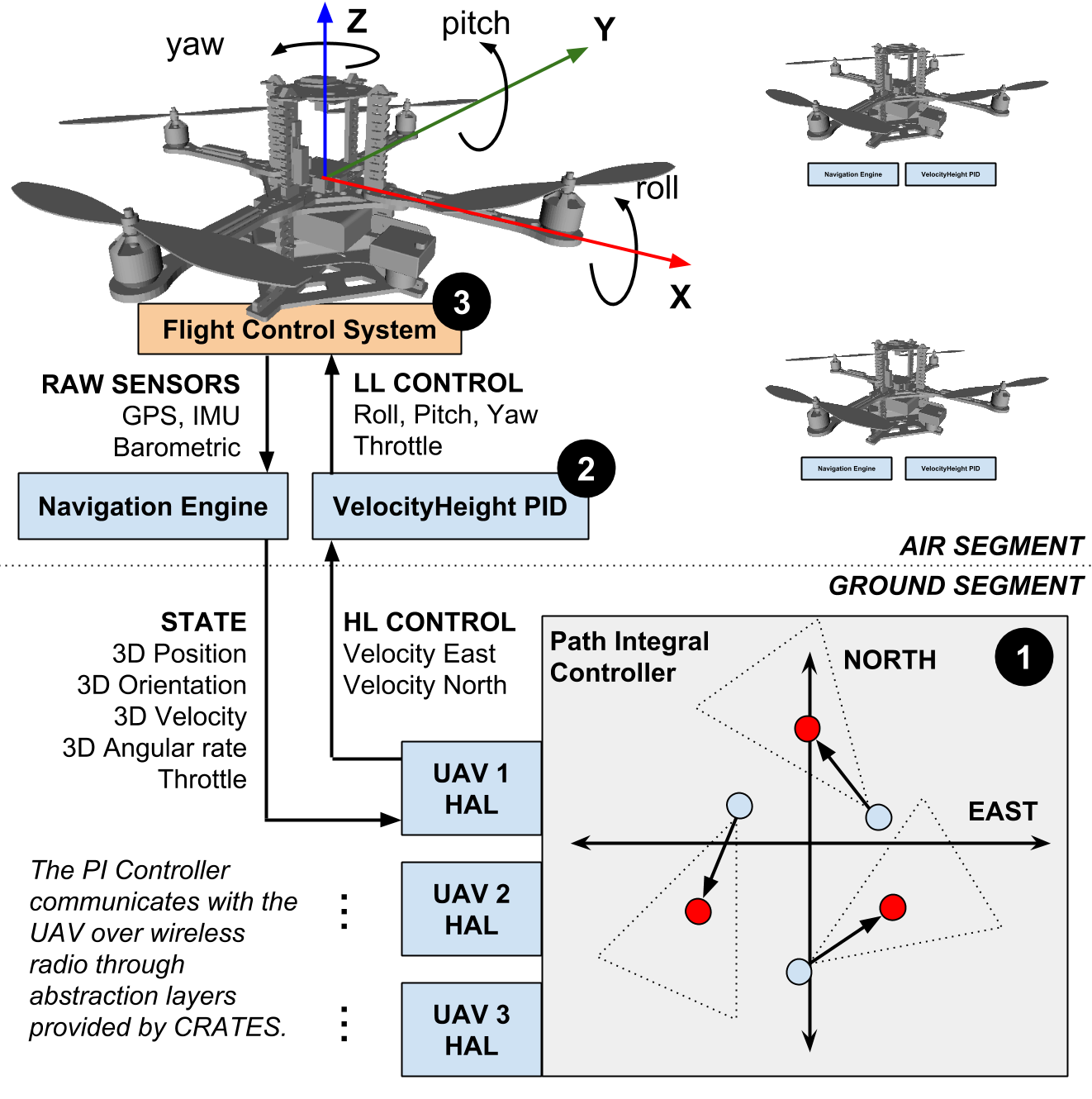}\caption{\textbf{Control hierarchy}: The path-integral controller (1) calculates target
velocities/heights for each quadrotor. These are converted to roll,
pitch, throttle and yaw rates by a platform-specific Velocity Height
PID controller (2). This control is in turn passed to the platform's
flight control system (3), and converted to relative motor speed changes.\label{fig:overview}}
\end{figure}
Quadrotors are a natural choice for an experimental platform, as they
provide a safe, highly-agile and inexpensive means by which to evaluate
UAV controllers. Figure~\ref{fig:overview} shows a 3D model of one
such quadrotor, the \emph{Ascending Technologies Pelican}. Quadrotors
have non-linear dynamics and are naturally unstable, making control
a non-trivial problem. 

Stochastic optimal control (SOC) provides a promising theoretical framework for achieving autonomous control of quadrotor systems. In contrast to deterministic control, SOC directly captures the uncertainty typically present in noisy environments and leads to solutions  that qualitatively depend on the level of uncertainty \cite{kappen2005path}.
However, with the exception of the simple Linear Quadratic Gaussian case, for which a closed form solution exists,  solving the SOC problem requires solving the Hamilton Jacobi Bellman (HJB) equations. These equations are generally intractable, and so the SOC problem remains an open challenge.

In such a complex setting, a hierarchical approach is usually taken and the control problem is reduced to follow a state-trajectory (or a set of way points) designed by hand or computed offline using trajectory planning algorithms \cite{Kendoulsurvey}. While the planning step typically involves a low-dimensional state representation, the control methods use a detailed complex state representation of the UAV.
Examples of control methods for trajectory tracking are the Proportional Integral Derivative or the Linear-Quadratic regulator.

A generic class of SOC problems was introduced in \citet{kappen2005path,NIPS2006_691} for which the controls and the cost function are restricted in a way that makes the HJB equation linear and therefore more efficiently solvable. This class of problems is known as path integral (PI) control, linearly-solvable controlled diffusions or Kullback-Leibler control, and it has lead to successful robotic applications, e.g.~\cite{KinjoFN2013,rombokas2012tendon,theu_jmlr}. A particularly interesting feature of this class of problems is that the computation of optimal control is an inference problem with a solution given in terms of the passive dynamics.
In a multi-agent system, where the agents follow independent passive dynamics, such a feature can be exploited using approximate inference methods such as variational approximations or belief propagation \cite{klinf,van2008graphical}.

In this paper, we show how PI control can be used for solving motion planning tasks on a team of quadrotors in real time. We combine periodic re-planning with receding horizon, similarly to model predictive control, with efficient importance sampling. At a high level, each quadrotor is modelled as a point mass that follows simple double integrator dynamics. Low-level control is achieved using a standard Proportional Integral Derivative (PID) velocity controller that interacts with a real or simulated flight control system. With this strategy we can scale PI control to ten units in simulation.
Although in principle there are no further limits to experiments with actual platforms, our first results with real quadrotors only include three units. To the best of our knowledge this has been the first real-time implementation of PI control on an actual multi-agent system.

In the next section we describe related work. We introduce our approach in Section~\ref{sec:method}~~~Results are shown on three different scenarios in Section \ref{sec:res}~ ~Finally, Section \ref{sec:conc}~ ~concludes this paper.

\section{Related Work on UAV Planning and Control}
\label{sec:rw}
There is a large and growing body of literature related to this topic. In this section, we highlight some of the most related papers to the presented approach. An recent survey of control methods for general UAVs can be found in~\citet{Kendoulsurvey}.

Stochastic optimal control is mostly used for UAV control in its simplest form, assuming a linear model perturbed by additive Gaussian noise and subject to quadratic costs (LQG), e.g. \cite{how2008}. While LQG can successfully perform simple actions like hovering, executing more complex actions requires considering additional corrections for aerodynamic effects such as induced power or blade flapping \cite{Hoffmann20111023}.  These approaches are mainly designed for accurate trajectory control and assume a given desired state trajectory that the controller transforms into motor commands.

Model Predictive Control (MPC) has been used optimize trajectories in multi-agent UAV systems \cite{shim2003decentralized}. MPC employs a model of the UAV and solves an optimal control problem at time $t$ and state $x(t)$ over a future horizon of a fixed number of time-steps. The first optimal move $u^*(t)$ is then applied and the rest of the optimal sequence is discarded. The process is repeated again at time $t+1$. A quadratic cost function is typically used, but other more complex functions exist.

MPC has mostly been used in indoor scenarios, where high-precision motion capture systems are available. For instance, in \citet{Kushleyev2013} authors generate smooth trajectories through known 3-D environments satisfying specifications on intermediate waypoints and show remarkable success controlling a team of $20$ quadrotors. 
Trajectory optimization is translated to a relaxation of a mixed integer quadratic program problem with additional constraints for collision avoidance, that can be solved efficiently in real-time.
Examples that follow a similar methodology can be found in~\citett{Turpin2012,Augugliaro12}\mbox{}. Similarly to our approach, these methods use a simplified model of dynamics, either using the 3-D position and yaw angle \citet{Kushleyev2013,Turpin2012}
or the position and velocities as in \citet{Augugliaro12}. However, these approaches are inherently deterministic and express the optimal control problem as a quadratic problem. In our case, we solve an inference problem by sampling and we do not require intermediate trajectory waypoints.

In outdoor conditions, motion capture is difficult and Global Positioning System (GPS) is used instead. Existing control approaches are typically either based on Reynolds flocking \cite{Burkle2011,Hauert2011,vasarhelyi2014outdoor,Reynolds} or flight formation \cite{guerrero2012flight,formationCCC}. In Reynolds flocking, each agent is considered a point mass that obeys simple and distributed rules: separate from neighbors, align with the average heading of neighbors and steer towards neighborhood centroid to keep cohesion. Flight formation control is typically modeled using graphs, where every node is an agent that can exchange information with all or several agents. Velocity and/or position coordination is usually achieved using consensus algorithms.

The work in \citet{Quintero2013} shares many similarities with our approach. Authors derive a stochastic optimal control formulation of the flocking problem for fixed-wings UAVs. They take a leader-follower strategy, where the leader follows an arbitrary (predefined) policy that is learned offline and define the immediate cost as a function of the distance and heading with respect to the leader. Their method is demonstrated outdoors with $3$ fixed-wing UAVs in a distributed sensing task. As in this paper, they formulate a SOC problem and perform MPC. However, in our case we do not restrict to a leader-follower setup and consider a more general class of SOC problems which can include coordination and cooperation problems.

\subsubsection{Planning approaches}

Within the planning community, \citet{bernardini2014planning} consider search and tracking tasks, similar to one of our scenarios.
Their approach is different to ours, they formulate a planning problem that uses used \emph{search patterns} that must be selected and sequenced to maximise the probability of rediscovering the target.
\citet{albore2015online} and \citet{chanel2013} consider a different problem: dynamic data acquisition and environmental knowledge optimisation.
Both techniques use some form of replanning. While \citet{albore2015online} uses a Markov Random Field framework to represent knowledge about the uncertain map and its quality, 
\citet{chanel2013} rely on partially-observable MDPs.
All these works consider a single UAV scenario and low-level control is either neglected or deferred to a PID or waypoint controller.

\subsubsection{Recent Progress in Path-Integral Control}
There has been significant progress in PI control, both theoretically and in applications.
Most of existing methods use parametrized policies to overcome the main limitations (see Section~\ref{sec:pi}).
Examples can be found in \citet{theu_jmlr,stulpcov,gomez14}. In these methods, the optimal control solution
is restricted by the class of parametrized policies and, more importantly, it is computed offline.
In~\citet{Rawlik2013}, authors propose to approximate the transformed cost-to-go function using linear operators
in a reproducing kernel Hilbert space. Such an approach requires an analytical form of the PI embedding, which is difficult to obtain in general.
In~\citet{horowitz2014linear}, a low-rank tensor representation is used to represent the model dynamics, allowing to scale PI control up to a 12-dimensional system. 
More recently, the issue of state-dependence of the optimal control has been addressed~\cite{thijssen2014path}, where a parametrized state-dependent feedback controller is derived for the PI control class.

Finally, model predictive PI control has been recently proposed for controlling a nano-quadrotor in indoor settings in an obstacle avoidance task~\cite{grady}. In contrast to our approach, their method is not hierachical and uses naive sampling, which makes it less sample efficient.
Additionally, the control cost term is neglected, which can have important implications in complex tasks involving noise.
The approach presented here scales well to several UAVs in outdoor conditions and is illustrated in tasks beyond obstacle avoidance navigation.

\section{Path-Integral Control for Multi-UAV planning}
\label{sec:method}

We first briefly review PI control theory.
This is followed by a description of the proposed method used to achieve motion planning of multi-agent UAV systems using PI control.

\subsection{Path-Integral Control}
\label{sec:pi}
We consider continuous time stochastic control problems, where the dynamics and cost are respectively linear and quadratic in the control input, but arbitrary in the state. More precisely, consider the following stochastic differential equation of the state vector $\bx\in\mathbb{R}^n$ under controls $\bu\in\mathbb{R}^m$
\begin{align}
\label{eq:dynamics}
d\bx & = \bof(\bx)dt + \bG(\bx)(\bu dt + d\bxi), 
\end{align}
where $\bxi$ is $m-$dimensional Wiener noise with covariance $\cov\in\mathbb{R}^{m\times m}$ and $\bof(\bx)\in\mathbb{R}^n$ and $\bG(\bx)\in\mathbb{R}^{n\times m}$ are arbitrary functions, $\bof$ is the drift in the uncontrolled dynamics (including gravity, Coriolis and centripetal forces), and $\bG$ describes the effect of the control $\bu$ into the state vector $\bx$.

A realization $\btau = \bx_{0:dt:T}$ of the above equation is called a (random) path. In order to describe a control problem we define the cost that is attributed to a path (cost-to-go) by
\begin{align}\label{eq:cost}
S(\btau|\bx_0, \bu) &=  r_T(\bx_T) \notag\\ 
	+&\sum_{t = 0:dt:T-dt} \left(r_t(\bx_t) +
	\frac{1}{2}\bu_t^{\top}\bR\bu_t\right)dt,	
\end{align}
where $r_T(\bx_T)$ and $r_t(\bx_t)$ are arbitrary state cost terms at end and intermediate times, respectively.
$\bR$ is the control cost matrix. The general stochastic optimal control problem is to minimize the expected cost-to-go w.r.t. the control
\begin{align*}
& \bu^* = \arg\min_{\bu} \mathbb{E}[ S(\btau|\bx_0, \bu)].
\end{align*}

In general, such a minimization leads to the Hamilton-Jacobi-Bellman (HJB) equations, which are non-linear, second order partial differential equations. 
However, under the following relation between the control cost and noise covariance
$\cov = \lambda\bR^{-1}$, 
the resulting equation is \emph{linear} in the exponentially transformed cost-to-go function.
The solution is given by the Feynman-Kac Formula, which expresses optimal control in terms of a Path-Integral, which can be interpreted as taking the expectation under the optimal path distribution~\cite{kappen2005path}
\begin{align}
p^*(\btau|\bx_0) 
	& \propto p(\btau|\bx_0, \bu) \exp(-S(\btau|\bx_0, \bu)/\lambda), \label{eq:p*} \\
\left<\bu^*_t(\bx_0)\right> 
	& = \left<\bu_t + (\bxi_{t + dt} - \bxi_{t})/dt\right>,
	  \label{eq:u*pi} 
\end{align}
where $p(\btau|\bx_0, \bu)$ denotes the probability of a (sub-optimal) path under equation~\eqref{eq:dynamics} and  $\left<\cdot\right>$ denotes expectation over paths distributed by $p^*$. 

The constraint $\cov = \lambda\bR^{-1}$ forces control and noise to act in the same dimensions, but in an inverse relation. Thus, for fixed $\lambda$, the larger the noise, the cheaper the control and vice-versa. Parameter $\lambda$ act as a temperature: higher values of $\lambda$ result in optimal solutions that are closer to the uncontrolled process.

Equation~\eqref{eq:u*pi} permits optimal control to be calculated by probabilistic inference methods, e.g., Monte Carlo. An interesting fact is that equations~(\ref{eq:p*}, \ref{eq:u*pi}) hold for all controls $\bu$. In particular, $\bu$ can be chosen to reduce the variance in the Monte Carlo computation of $\left<\bu^*_t(\bx_0)\right> $ which amounts to importance sampling.
This technique can drastically improve the sampling efficiency, which is crucial in high dimensional systems. 
Despite this improvement, direct application of PI control into real systems is limited because
it is not clear how to choose a proper importance sampling distribution.
Furthermore, note that equation~\eqref{eq:u*pi} yields the optimal control for all times $t$ averaged over states. The result is therefore an open-loop controller that neglects the state-dependence of the control beyond the initial state.

\subsection{Multi-UAV planning}
\label{sec:piUAV} 
The proposed architecture is composed of two main levels.
At the most abstract level, the UAV is modeled as a 2D point-mass system that follows double integrator dynamics.
At the low-level, we use a detailed second order model that we learn from real flight data~\cite{denardi2008}.
We use model predictive control combined with importance sampling.
There are two main benefits of using the proposed approach: first, since the state is continuously updated, the controller does not suffer from the problems caused by using an open-loop controller. Second, the control policy is not restricted by any parametrization. 

The two-level approach permits to transmit control signals from the high-level PI controller to the low-level control system at a relatively low frequencies (we use $15$Hz in this work).
Consequently, the PI controller has more time available for sampling a large number of trajectories, which is critical to obtain good estimates of the
control.
The choice of 2D in the presented method is not a fundamental limitation, as long as double-integrator dynamics is used. The control hierarchy introduces additional model mismatch. However, as we show in the results later, this mismatch is not critical for obtaining good performance in real conditions.

Ignoring height, the state vector $\bx$ is thus composed of the East-North (EN) positions and EN velocities of each agent $i = 1, \ldots, M$ as $x_i = [p_i, v_i]^\top$ where $p_i, v_i\in\mathbb{R}^2$. Similarly, the control $\bu$ consists of EN accelerations $u_i\in\mathbb{R}^2$. Equation~\eqref{eq:dynamics} decouples between the agents and takes the linear form
\begin{align}
dx_i & = Ax_i dt + B(u_idt + d\xi_i),\notag\\
A    & = \left[\begin{array}{cc}
                0 & 1 \\ 0 & 0\end{array}\right], \qquad B= \left[\begin{array}{c}0 \\ 1\end{array}\right]. \label{eq:ab}
\end{align}
Notice that although the dynamics is decoupled and linear, the state cost $r_t(\bx_t)$ in equation~\eqref{eq:cost} can be any arbitrary function of all UAVs states. As a result, the optimal control will in general be a non-linear function that couples all the states and thus hard to compute.

\begin{algorithm}[t]
\begin{algorithmic}[1]
\caption{PI control for UAV motion planning}\label{algo}
\Function{PIController}{$N,H,dt,r_t(\cdot),\cov,\bu_{t:dt:t + H}$}
\For{$k = 1, \ldots, N$}
\State Sample paths $\btau_k = \{ \bx_{t:dt:t+H}\}_k$ with Eq.~\eqref{eq:ab}
\EndFor
\State Compute $S_k = S(\btau_k| \bx_0, \bu)$ with Eq.~\eqref{eq:cost}
\State Store the noise realizations $\{\bxi_{t:dt:t+H}\}_k$
\State Compute the weights: $w_k = e^{-S_k/\lambda}/\sum_l e^{-S_l/\lambda}$
\For{$s = t:dt:t+H$}
\State $\bu^*_s = \bu_s + \frac1{dt}\sum_k w_k \left(\{\bxi_{s + dt}\}_k - \{\bxi_{s}\}_k\right)$
\EndFor
\State \textbf{Return} next desired velocity: $\bv_{t + dt} = \bv_{t} + \bu^*_t dt$ and $\bu^*_{t:dt:t+H}$ for importance sampling at $t+dt$
\EndFunction
\end{algorithmic}
\end{algorithm}

Given the current joint optimal action $\bu^*_t$ and velocity $\bv_t$, the expected velocity at the next time $t'$ is 
calculated as $\bv_{t'} = \bv_t + (t' - t) \bu^*_t $ and passed to the low-level controller. The final algorithm optionally keeps an importance-control sequence $\bu_{t:dt:t + H}$ that is incrementally updated. We summarize the high-level controller in Algorithm~\ref{algo}.

The importance-control sequence $\bu_{t:dt:t + H}$ is initialized using prior knowledge or with zeros otherwise. Noise is dimension-independent, i.e.~$\cov = \sigma_u^2\text{Id}$.
To measure sampling convergence, we define the \emph{Effective Sample Size} (ESS) as
$\text{ESS} := 1/\sum_{k=1}^N w_k^2$, which is a quantity between $1$ and $N$. Values of ESS close to one indicate an estimate dominated by just one sample and a poor estimate of the optimal control, whereas an ESS close to $N$ indicates near perfect sampling, which occurs when the importance- equals the optimal-control function.

\subsection{Low Level Control}
The target velocity $\bv=\left[v_{E}\: v_{N}\right]^{\top}$ is passed
along with a height $\hat{p}_{U}$ to a Velocity-Height controller.
This controller uses the current state estimate
of the real quadrotor $\mathbf{y}=\left[p_{E}\: p_{N}\: p_{U}\:\phi\:\theta\:\psi\: u\: v\: w\: p\: q\: r\right]^{\top}$,
where  $(p_{E},p_{N},p_{U})$ and $(\phi,\theta,\psi)$ denote navigation-frame position and orientation
and $(u,v,w),(p,q,r)$ denote body-frame and angular velocities, respectively.
It is composed of four independent PID controllers for roll $\hat{\phi}$,
pitch $\hat{\theta}$, throttle $\hat{\gamma}$ and yaw rate $\hat{r}$.
that send the commands to the flight control system (FCS) to achieve $\bv$.
\begin{figure}[!t]
\centering{}\includegraphics[width=\columnwidth]{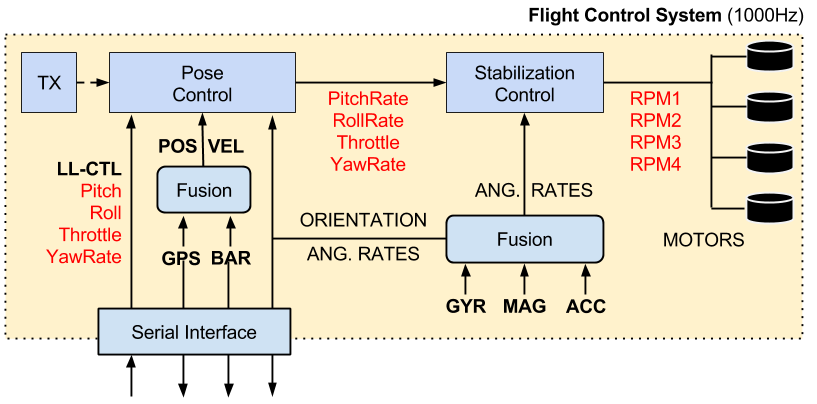}\caption{The flight control system (FCS) is comprised of two control loops: one for stabilization and the other for pose control. A low-level controller interacts with the FCS over a serial interface to stream measurements and issue control.\label{fig:fcs}}
\end{figure}

Figure~\ref{fig:fcs} shows the details of the FCS.
The control loop runs at 1kHz fusing triaxial gyroscope, accelerometer and magnetometer measurements. The accelerometer and magnetometer measurements are used to determine a reference global orientation, which is in turn used to track the gyroscope bias.
The difference between the desired and actual angular rates are converted to motor speeds using the model in \citet{Mahony2012}.



An outer pose control loop calculates the desired angular rates based
on the desired state. 
Orientation is obtained from the inner control loop, while position
and velocity are obtained by fusing GPS
navigation fixes with barometric pressure (BAR) based altitude measurements.
The radio transmitter (marked TX in the diagram) allows the operator
to switch quickly between autonomous and manual control of a platform.
 There is also an acoustic alarm on the platform
itself, which warns the operator when the GPS signal is lost or the
battery is getting low. If the battery reaches a critical level or
communication with the transmitter is lost, the platform can be configured
to land immediately or alternatively, to fly back and land at its
take-off point.


\subsection{Simulator Platform}
We have developed an open-source framework
called \sname \footnote{\sname stands for 'Cognitive Robotics Architecture for Tightly-Coupled Experiments and Simulation'. Available at \url{https://bitbucket.org/vicengomez/crates}}.
The framework is a implementation of \qrsim \cite{qrsim,iros_crates} in Gazebo, which
uses Robot Operating System (ROS) for high-level control.
It permits high-level controllers to be platform-agnostic.
It is similar to the Hector Quadrotor project \cite{meyer2012} with a formalized notion of a
hardware abstraction layers. 

The \sname simulator propagates the quadrotor state forward in time
based on a second order model \cite{denardi2008}. The equations
were learned from real flight data and verified by expert domain knowledge.
In addition to platform dynamics, \sname also simulates various noise-perturbed
sensors, wind shear and turbulence. Orientation and barometric altitude
errors follow zero-mean Ornstein-Uhlenbeck processes, while GPS error
is modeled at the pseudo range level using trace data available from
the International GPS Service. %
In accordance with the Military Specification MIL-F-8785C,
wind shear is modeled as a function of altitude, while turbulence
is modeled as a discrete implementation of the Dryden
model. \sname also provides support for generating terrain
from satellite images and light detection and ranging (LIDAR) technology, and reporting collisions between platforms and terrain.

\section{Results}
\label{sec:res} 

\begin{figure}[t]
\centering{}\includegraphics[width=.48\columnwidth]{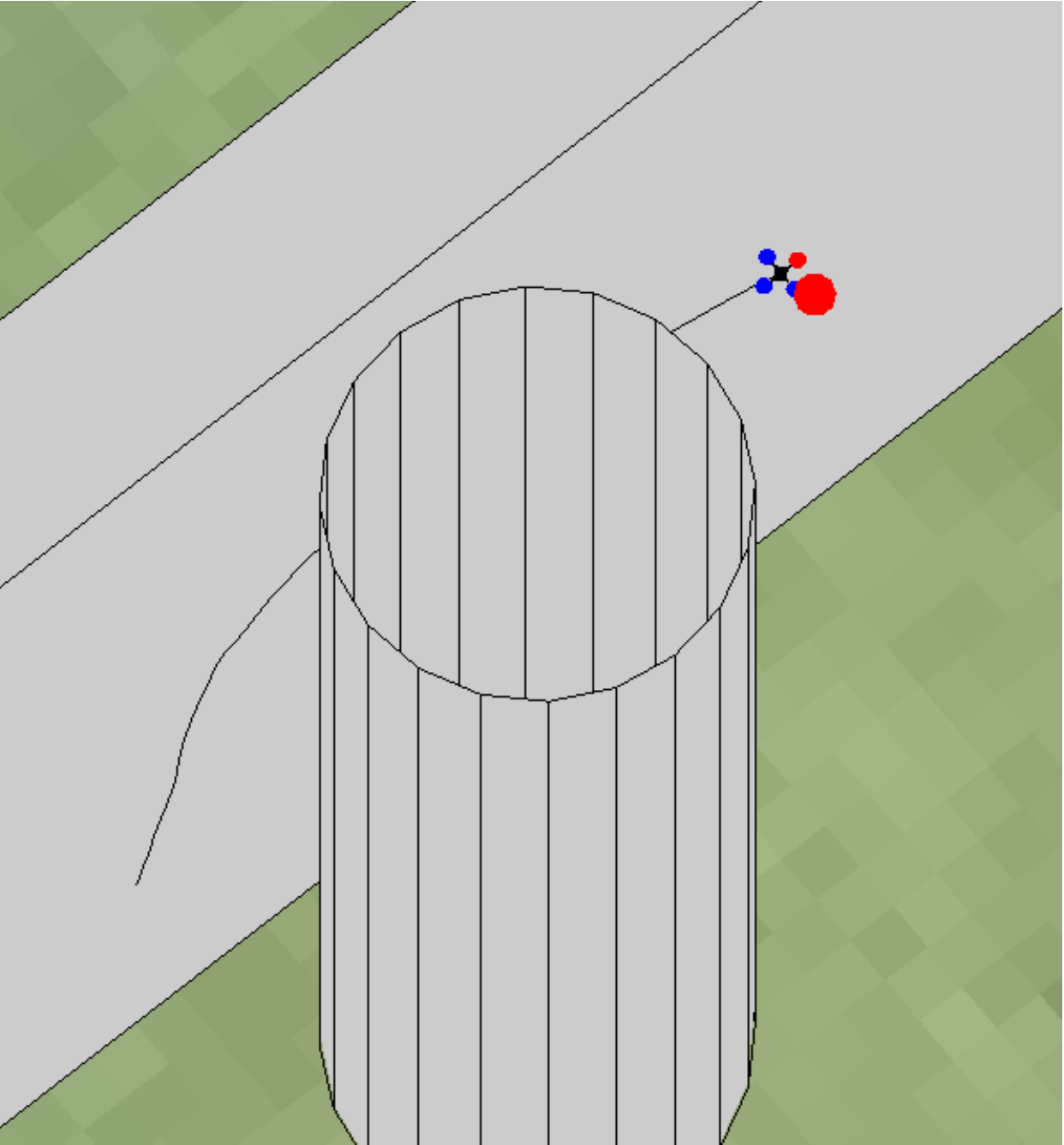}
\centering{}\includegraphics[width=.48\columnwidth]{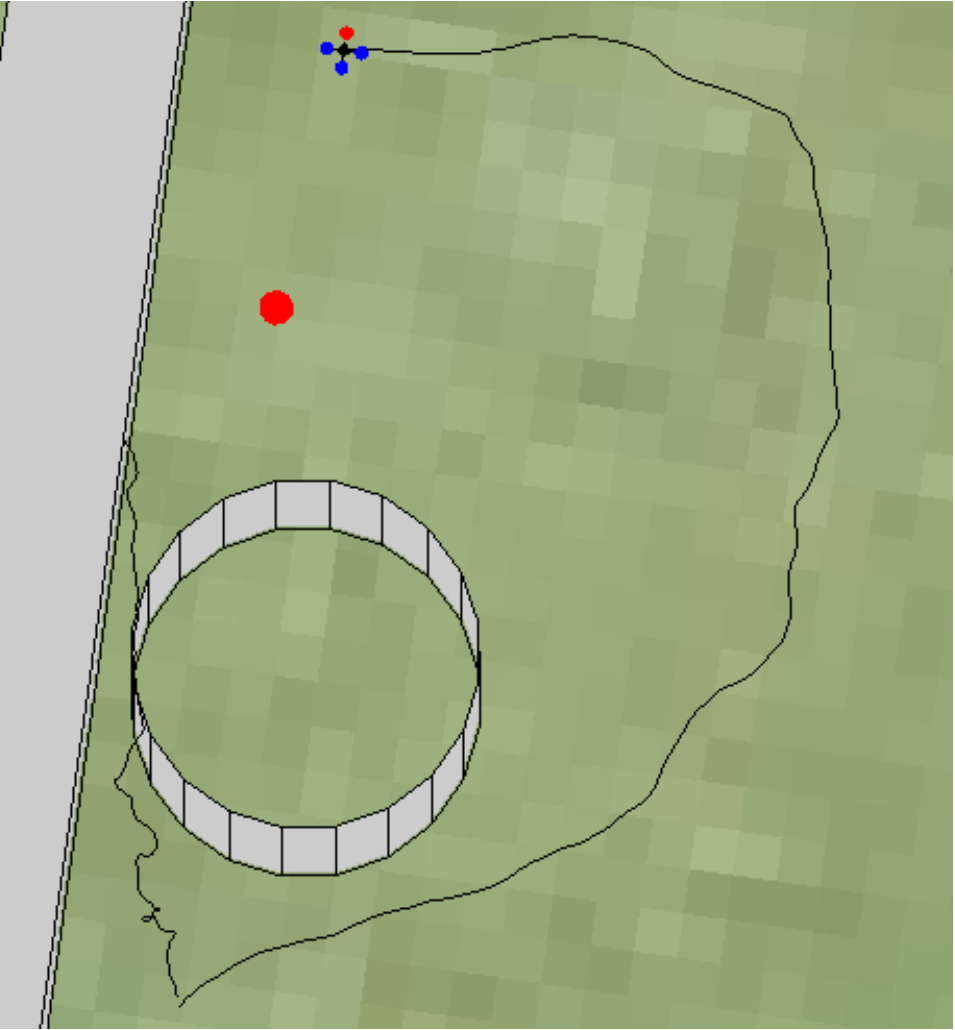}
\caption{Drunken Quadrotor: a red target has to be reached while avoiding obstacles. (Left) the shortest route is the optimal solution in the absence of noise. (Right) with control noise, the optimal solution is to fly around the building.\label{fig:avoid}}
\end{figure}

We now analyze the performance of the proposed approach in three different tasks.
We first show that, in the presence of control noise, PI control is preferable over other approaches.
For clarity, this scenario is presented for one agent only.
We then consider two tasks involving several units: a flight formation task and a 
pursuit-evasion task.

We compare the PI control method described in Section \ref{sec:piUAV} with iterative linear-quadratic Gaussian (iLQG) control \cite{Todorov05}.
iLQG is a state-of-the-art method based on differential dynamic programming, that iteratively computes local linear-quadratic approximations to the finite horizon problem.
A key difference between iLQG and PI control is that the linear-quadratic approximation is certainty equivalent. Consequently, iLQG yields a noise independent solution.

\subsection{Scenario I: Drunken Quadrotor}
This scenario is inspired in \citet{kappen2005path} and highlights the benefits of SOC in a quadrotor task. The Drunken Quadrotor is a finite horizon task where a quadrotor has to reach a target, while avoiding a building and a wall (figure~\ref{fig:avoid}). There are two possible routes: a shorter one that passes through a small gap between the wall and the building, and a longer one that goes around the building. Unlike SOC, the deterministic optimal solution does not depend on the noise level and will always take the shorter route. However, with added noise, the risk of collision increases and thus the optimal noisy control is to take the longer route.

\begin{figure*}[t]
\centering{}\includegraphics[width=.6\textwidth]{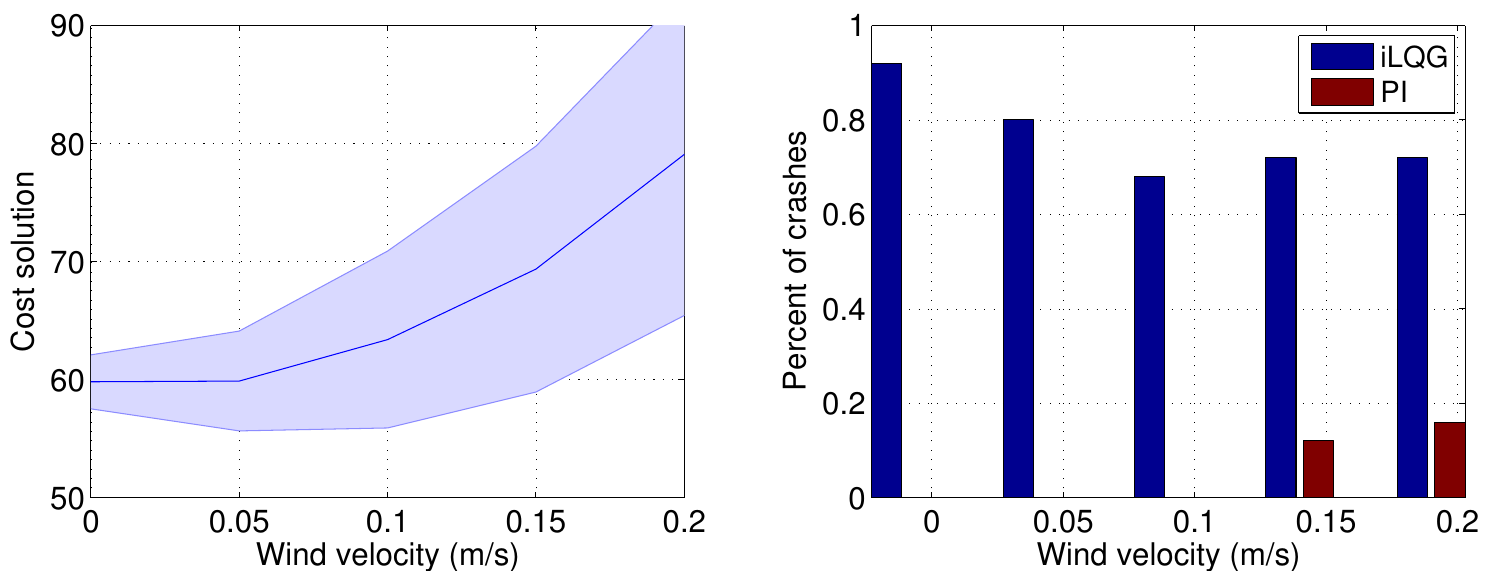}
\caption{Results: 
\textbf{Drunken Quadrotor with wind}:
For different wind velocities and fixed control noise $\sigma_u^2=0.5$. (Left)~cost of the obtained solutions and (Right) percentage of crashes using iLQG and PI.
 \label{fig:resultsA}}
\end{figure*}

This task can be alternatively addressed using other planning methods, such as the one proposed by~\citet{Ono}, which allow for specification of user's acceptable levels of risk using chance constraints.
Here we focus on comparing deterministic and stochastic optimal control for motion planning.
The amount of noise thus determines whether the optimal solution is go through the risky path or the longer safer path.

The state cost in this problem consists of hard constraints that assign infinite cost when either the wall or the building is hit. PI control deals with collisions by killing particles that hit the obstacles during Monte Carlo sampling. For iLQG, the local approximations require a twice differentiable cost function.
We resolved this issue by adding a smooth obstacle-proximity penalty in the cost function.
Although iLQG computes linear feedback, we tried to improve it with a MPC scheme, similar as for PI control. Unfortunately, this leads to numerical instabilities in this task, since the system disturbances tend to move the reference trajectory through a building when moving from one time step to the next.
For MPC with PI control we use a receding horizon of three seconds and perform re-planning at a frequency of $15$ Hz with $N = 2000$ sample paths.
Both methods are initialized with $\bu_t =0, \forall t$. 
iLQG requires approximately $10^3$ iterations to converge with a learning rate of $0.5\%$.

Figure~\ref{fig:avoid} (left) shows an example of real trajectory computed for low control noise level, $\sigma^2_u = 10^{-3}$.
To be able to obtain such a trajectory we deactivate sensor uncertainties in accelerometer, gyroscope, orientation and altimeter.
External noise is thus limited to aerodynamic turbulences only.
In this case, both iLQG and PI solutions correspond to the shortest path, i.e.~go through the gap between the wall and the building.
Figure~\ref{fig:avoid} (right) illustrates the solutions obtained for larger noise level $\sigma_u^2 = 1$.
While the optimal reference trajectory obtained by iLQG does not change, which results in collision once the real noisy controller is executed (left path), the PI control solution avoids the building and takes the longer route (right path). 
Note that iLQG can find both solutions depending on initialization. However, However, it will always choose the shortest route, regardless of nearby obstacles. 
Also, note that the PI controlled unit takes a longer route to reach the target. The reason is that the control cost $\bR$ is set quite high in order to reach a good ESS. Alternatively, if $\bR$ is decreased, the optimal solution could reach the target sooner, but at the cost of a decreased ESS. This trade-off, which is inherent in PI control, can be resolved by incorporating feedback control in the importance sampling, as presented in~\citet{thijssen2014path}. 

We also consider more realistic conditions with noise not limited to act in the control.
Figure~\ref{fig:resultsA}~(a,b) shows results in the presence of wind and sensor uncertainty.
Panel~(a) shows how the wind affects the quality of the solution, resulting in an increase of the variance and the cost for stronger wind.
In all our tests, iLQG is not able to bring the quadrotor to the other side.
Panel~(b) shows the percentage of crashes using both methods.
Crashes occur often using iLQG control and only occasionally using PI control. With stronger wind, the iLQG controlled unit does occasionally
not even reach the corridor (the unit did not reach the other side but did not crash either). This explains the difference in percentages of Panel~(b).
We conclude that for multi-modal tasks (tasks where multiple solution trajectories exist), the proposed method is preferable to iLQG.

\subsection{Scenario II: Holding Pattern}
The second scenario addresses the problem of coordinating agents to hold their position near a
point of interest while keeping a safe range of velocities and avoiding crashing into each other.
Such a problem arises for instance when multiple aircraft need to land at the same location, and simultaneous
landing is not possible.
The resulting flight formation has been used frequently in the literature \cite{vasarhelyi2014outdoor,how2008,formationCCC,franchi2012modeling},
but always with prior specification of the trajectories.
We show how this formation is obtained as the optimal solution of a SOC problem.
\begin{figure}[!ht]
\centering{}\includegraphics[width=.96\columnwidth]{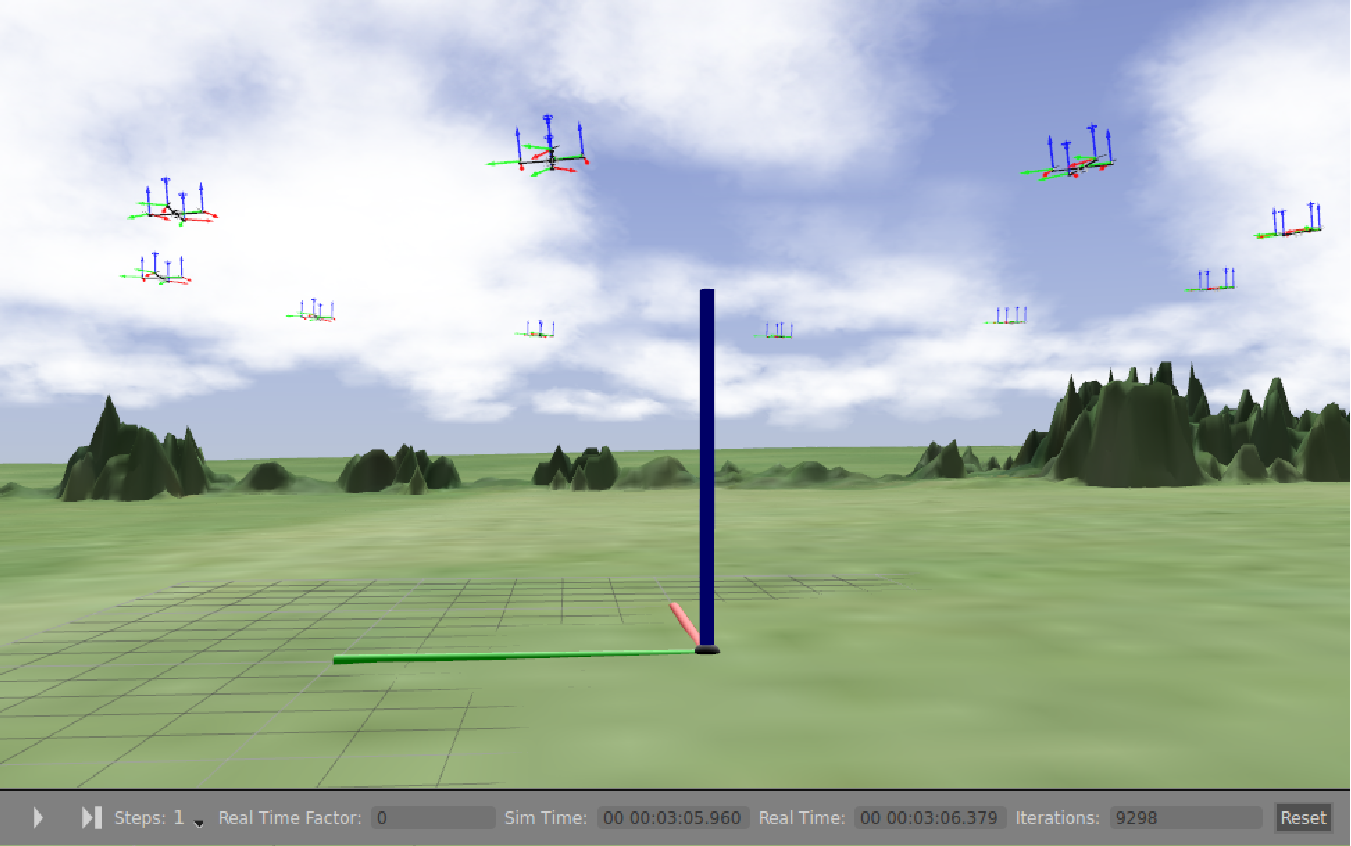}
\caption{
Holding pattern in the \sname simulator.
Ten units coordinate their flight in a circular formation.
In this example, $N=10^4$ samples, control noise is $\sigma_u^2=0.1$ and horizon $H=1$ sec.
Cost parameters are $v_{\text{min}}=1,v_{\text{max}}=3,C_{\text{hit}}=20$ and $\text{d}=7$.
Environmental noise and sensing uncertainties are modeled using realistic parameter values.
\label{fig:gazebohp}}
\end{figure}

Consider the following state cost (omitting time indexes)
\begin{align}\label{eq:hp}
r_\text{HP}(x)= & \sum_{i=1}^{M} \exp\left(v_i-v_{\text{max}}\right) + \exp\left( v_{\text{min}}-v_i\right)\notag\\
    & + \exp\left(\parallel p_i-\text{d}\parallel_2 \right) + \sum_{j>i}^{M}C_{\text{hit}}/ \parallel p_i-p_j\parallel_2
\end{align}
where $v_{\text{max}}$ and $v_{\text{min}}$ denote the maximum and minimum velocities, respectively,
$\text{d}$ denotes penalty for deviation from the origin and $C_{\text{hit}}$ is the penalty for collision risk of two agents.
$\parallel\cdot\parallel_2$ denotes $\ell$-$2$ norm.

The optimal solution for this problem is a circular flying pattern where units fly equidistantly from each other.
The value of parameter $\text{d}$ determines the radius and the average velocities of the agents are determined from
$v_{\text{min}}$ and $v_{\text{max}}$. Since the solution is symmetric with respect to the direction of rotation
(clockwise or anti-clockwise), only when the control is executed, a choice is made and the symmetry is broken.
Figure ~\ref{fig:gazebohp} shows a snapshot of a simulation after the flight formation has been reached
for a particular choice of parameter values
\footnote{Supplementary video material is available at \url{http://www.mbfys.ru.nl/staff/v.gomez/uav.html}}.
Since we use an uninformed initial control trajectory, there is a transient period during which the agents organize to reach the optimal configuration.
The coordinated circular pattern is obtained regardless of the initial positions.
This behavior is robust and obtained for a large range of parameter values. 

\begin{figure*}[t]
\centering{}\includegraphics[width=\textwidth]{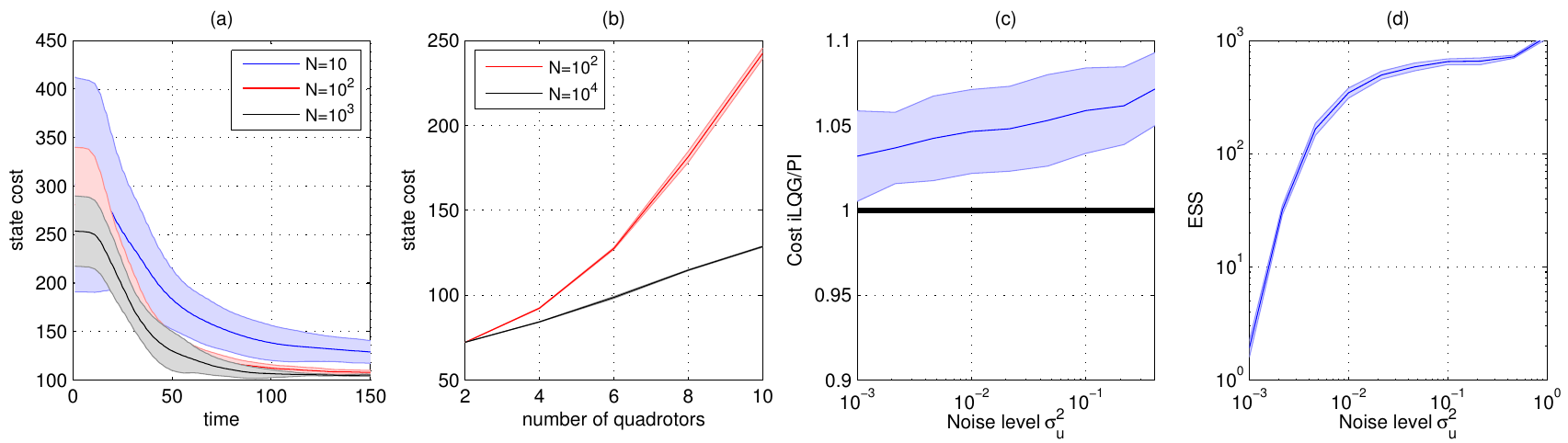}
\caption{\textbf{Holding pattern}:
(a)~evolution of the state cost for different number of samples $N=10,10^2,10^3$.
(b)~scaling of the method with the number of agents.
For different control noise levels, (c)~comparison between iLQG and PI control (ratios~$>1$ indicate better performance of PI over iLQG)
and (d)~Effective Sample Sizes.
Errors bars correspond to ten different random realizations.
 \label{fig:resultsB}}
\end{figure*}

Figure~\ref{fig:resultsB}(a) shows immediate costs at different times. 
Cost always decreases from the starting configuration until the formation is reached.
This value depends on several parameters.
We report its dependence on the number $N$ of sample paths.
For large $N$, the variances are small and the cost attains small values at convergence.
Conversely, for small $N$, there is larger variance and the obtained dynamical configuration
is less optimal (typically the distances between the agents are not the same). 
During the formation of the pattern the controls are more expensive.
For this particular task, full convergence of the path integrals is not required, and the formation can be achieved with a very small $N$.

Figure~\ref{fig:resultsB}(b) illustrates how the method scales as the number of agents increases.
We report averages over the mean costs over $20$ time-steps after one minute of flight.
We varied $M$ while fixing the rest of the parameters (the distance $\text{d}$ which was set equal to the number of agents in meters).
The small variance of the cost indicates that a stable formation is reached in all the cases.
As expected, larger values of $N$ lead to smaller state cost configurations.
For more than ten UAVs, the simulator starts to have problems in this task and occasional crashes may occur before the formation is reached due to limited sample sizes.
This limitation can be addressed, for example, by using more processing power and parallelization and it is left for future work.
\begin{figure}[!t]
\centering\includegraphics[width=.7\columnwidth]{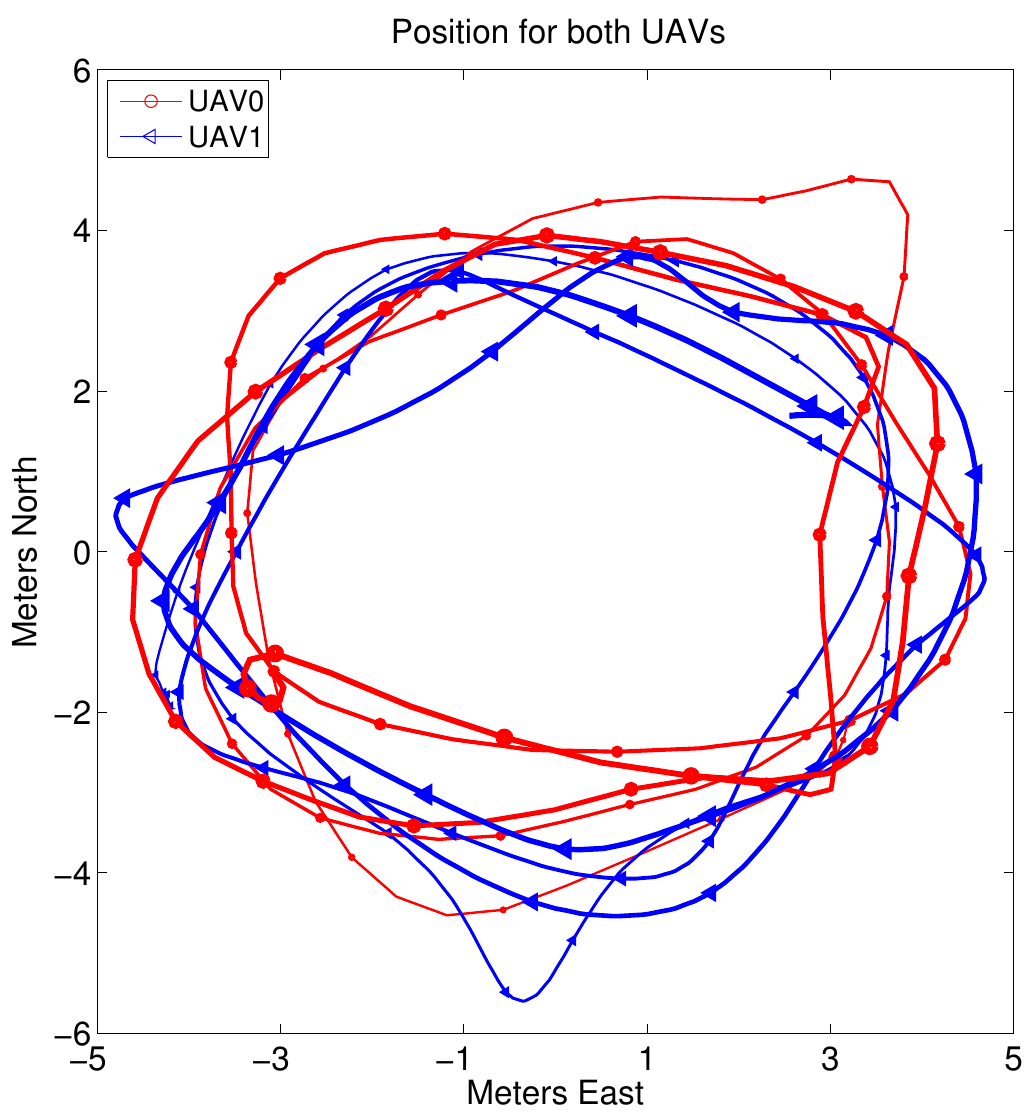}
\caption{Resulting trajectories of a Holding Pattern experiment using two platforms in outdoors conditions.
\label{fig:expresults}}
\end{figure}

\begin{figure*}[t]
\centering{}\includegraphics[width=.7\columnwidth]{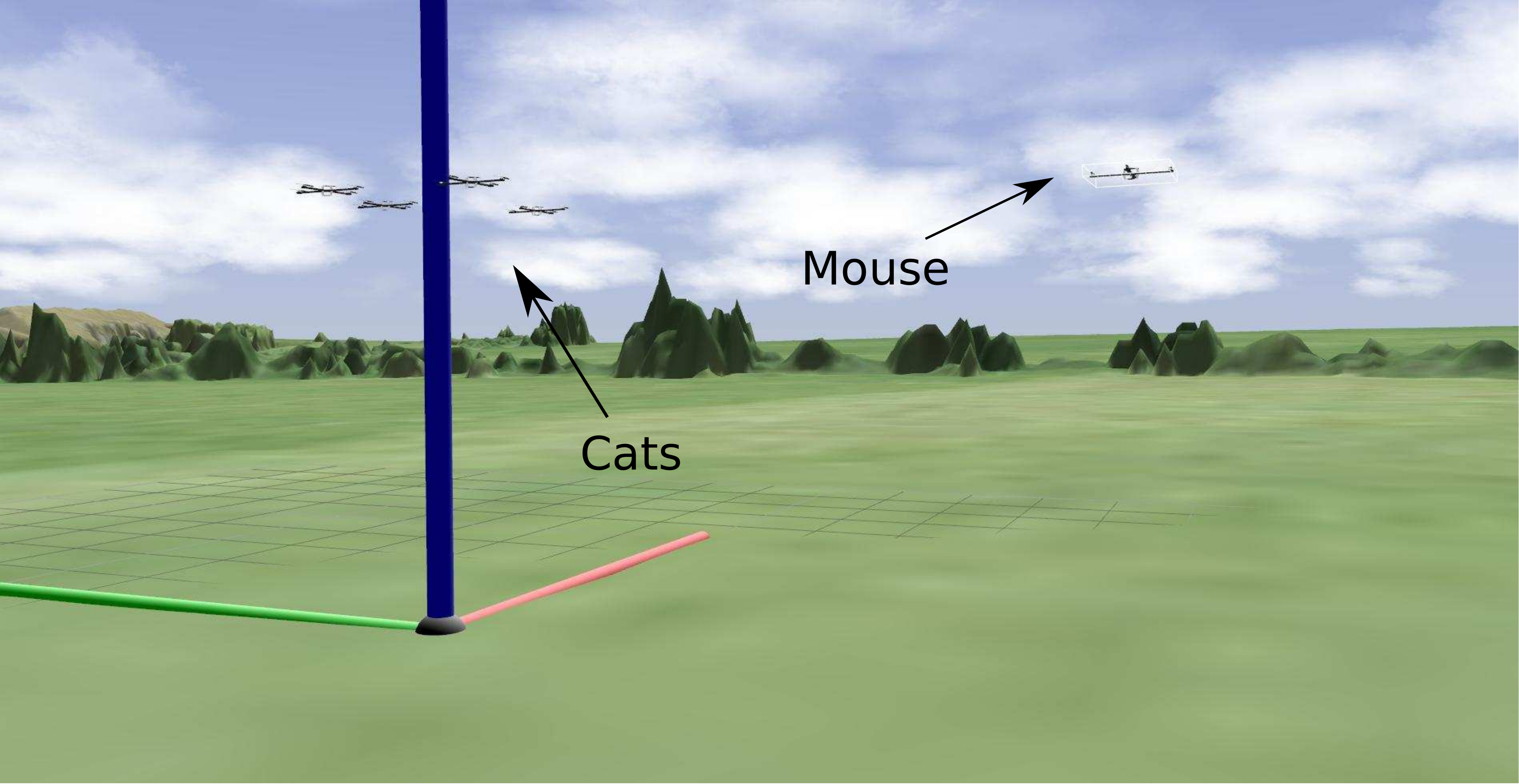}
\centering{}\includegraphics[width=.7\columnwidth]{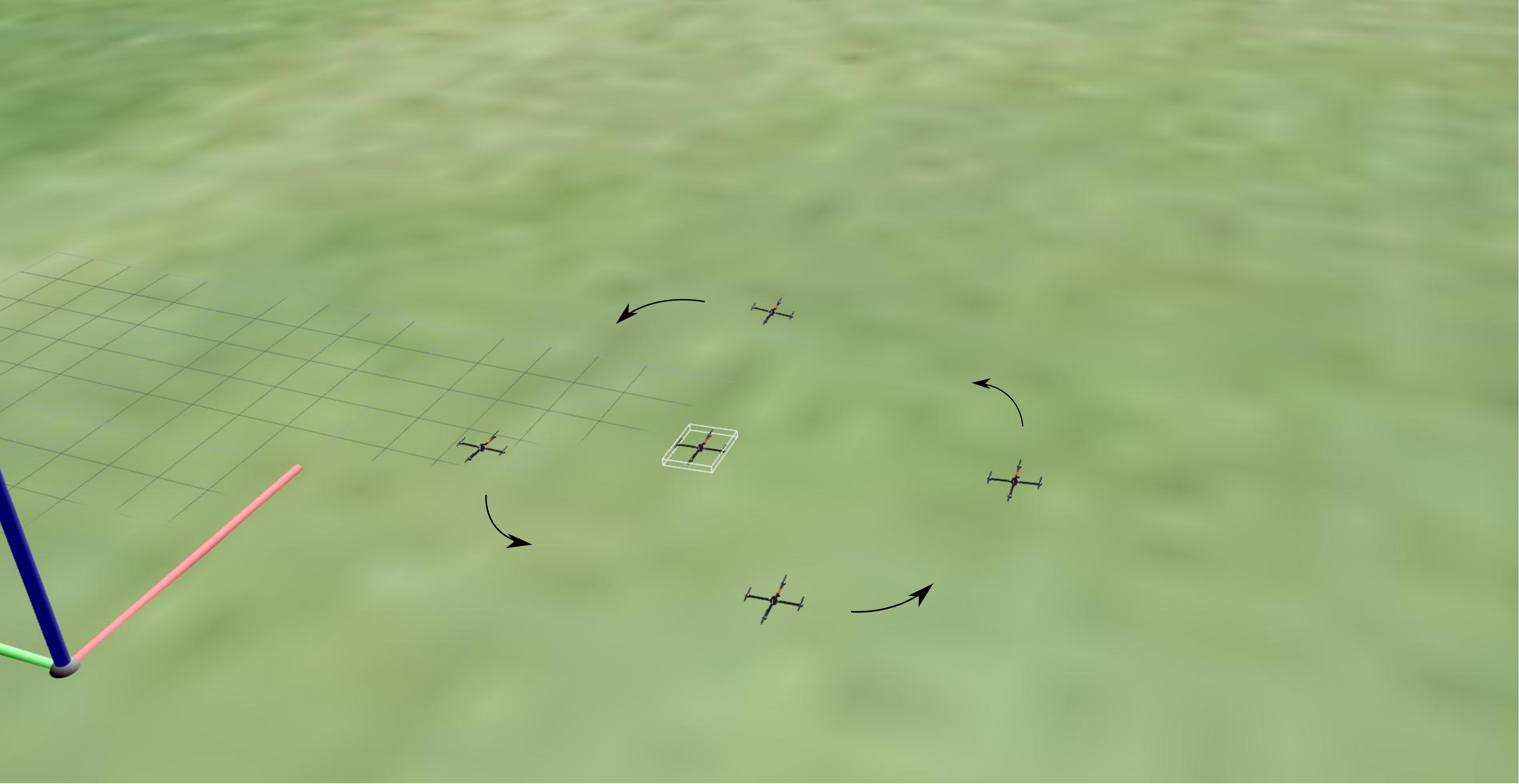}
\centering{}\includegraphics[width=.7\columnwidth]{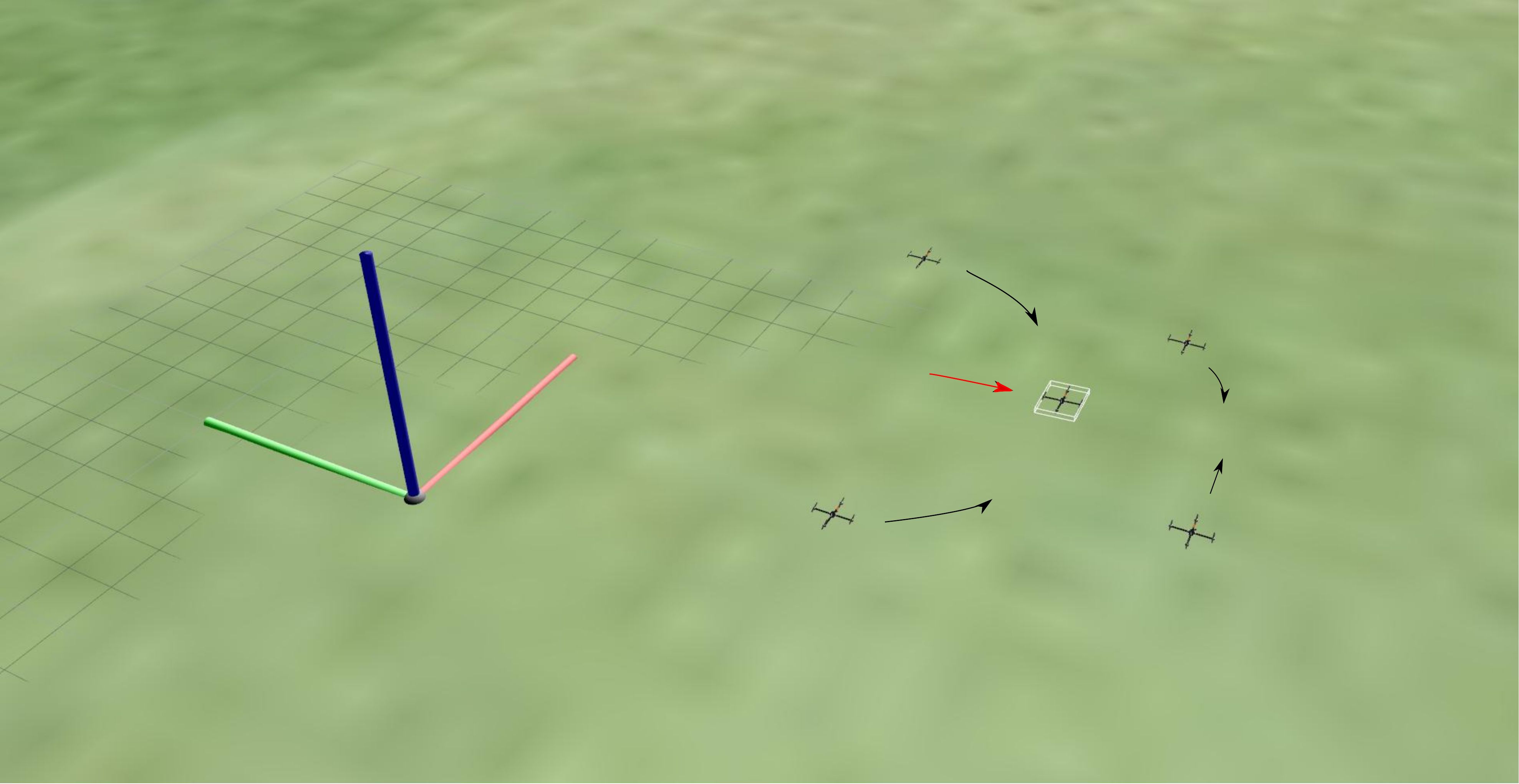}
\centering{}\includegraphics[width=.7\columnwidth]{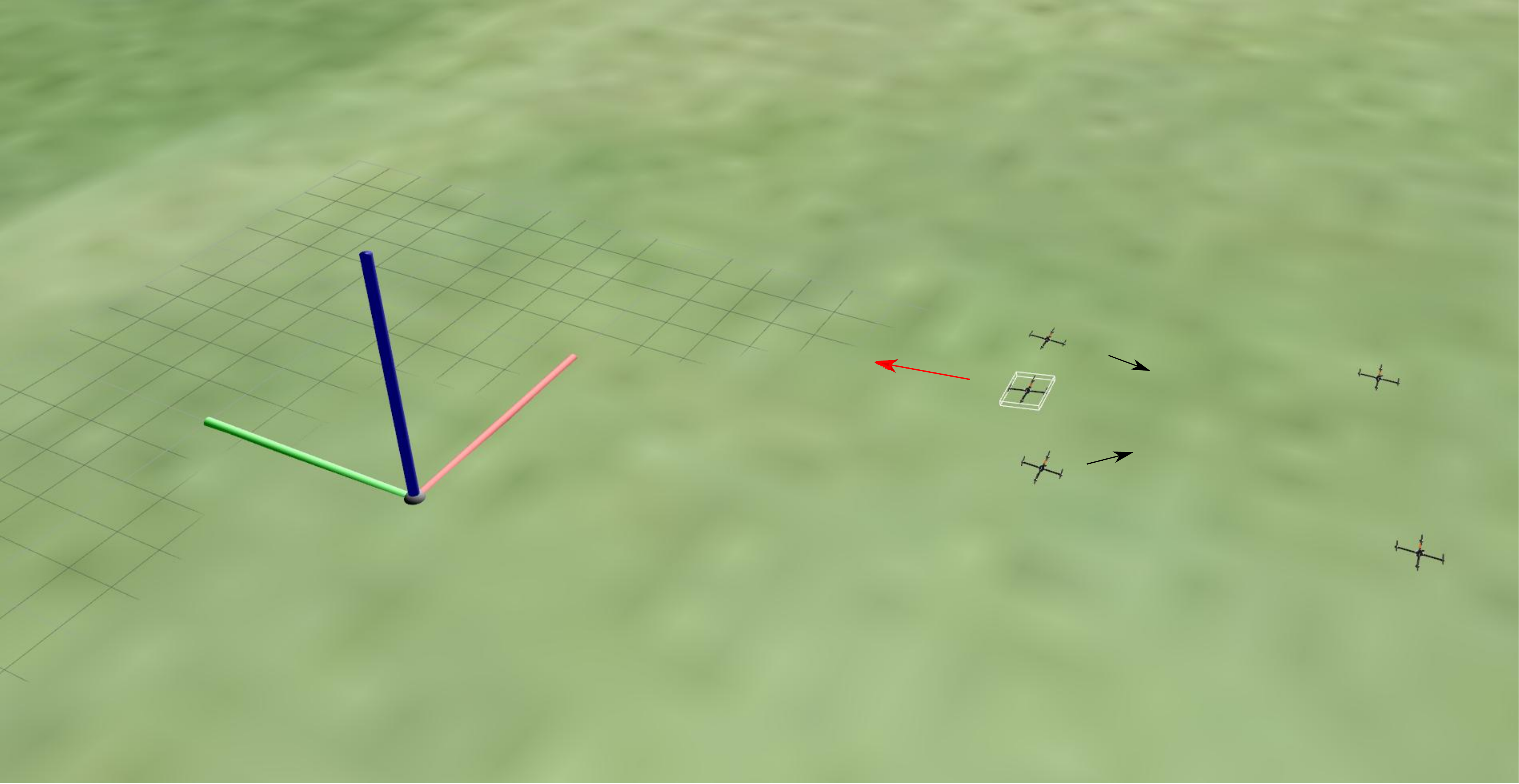}
\caption{
Cat and mouse scenario: \textbf{(Top-left)} four cats and one mouse.
{\bf(Top-right)} for horizon time $H=2$ seconds, the four cats surround the mouse forever and keep rotation around it.
{\bf(Bottom-left)} for horizon time $H=1$ seconds, the four cats chase the mouse but
{\bf(bottom-right)} the mouse manages to escape.
With these settings, the multi-agent system alternates between these two dynamical states.
Number of sample paths is $N=10^4$, noise level $\sigma^2_u=0.5$. Other parameter values are
$\text{d}=30, v_\text{min}=1, v_\text{max}=4, v_\text{min}=4$ and $v_{\text{max mouse}}=3$.
\label{fig:catmouse}}
\end{figure*}

We also compared our approach with iLQG in this scenario.
Figure~\ref{fig:resultsB}(c) shows the ratio of cost differences after convergence of both solutions.
 Both use MPC, with a horizon of 2s and update frequency of 15Hz.
Values above $1$ indicate that PI control consistently outperforms iLQG in this problem.
Before convergence, we also found, as in the previous task, that iLQG resulted in occasional crashes
while PI control did not.
The Effective Sample Size (ESS) is shown in Figure~\ref{fig:resultsB}(d). We observe
that higher control noise levels result in better exploration and thus better controls.
We can thus conclude that the proposed methodology is feasible for coordinating a large team of quadrotors.

For this task, we performed experiments with the real platforms.
Figure~\ref{fig:expresults} shows real trajectories obtained in outdoor conditions
(see also the video that accompanies this paper for an experiment with three platforms).
Despite the presence of significant noise, the circular behavior was also obtained.
In the real experiments, we used a Core i7 laptop with 8GB RAM as base station, which run its own ROS messaging core and forwarded messages to and
from the platforms over a IEEE 802.11 2.4GHz network. For safety reasons, the quadrotors were flown at different altitudes.

\subsection{Scenario III: Cat and Mouse}

The final scenario that we consider is the cat and mouse scenario.
In this task, a team of $M$ quadrotors (the cats) has to catch (get close to) another quadrotor (the mouse).
The mouse has autonomous dynamics: it tries to escape the cats by moving at velocity inversely proportional to the distance 
to the cats. More precisely, let $p_\text{mouse}$ denote the $2$D position of the mouse, the velocity command for the mouse is computed (omitting time indexes) as
\begin{align*}
v_{\text{mouse}} &= v_{\substack{\text{max} \\\text{mouse}}}
\frac{v}{\parallel v \parallel_2}, & \text{where } & v = \sum_{i=1}^{M}\frac{p_i-p_\text{mouse}}{\parallel p_i-p_\text{mouse}\parallel^2}.
\end{align*}
The parameter $v_{\substack{\text{max} \\\text{mouse}}}$ determines the maximum velocity of the mouse.
As state cost function we use equation~\eqref{eq:hp} with an additional penalty term that depends on the sum of the distances to the mouse
$$r_{\text{CM}}(x) = r_{\text{HP}}(x) + \sum_{i=1}^M  \parallel p_i-p_{\text{mouse}}\parallel_2.$$

This scenario leads to several interesting dynamical states.
For example, for a sufficiently large value of $M$, the mouse always gets caught (if its initial position is not close to the boundary,
determined by $\text{d}$).
The optimal control for the cats consists in surrounding the mouse to prevent collision. Once the mouse is surrounded, the cats keep rotating around it, as in the
previous scenario, but with the origin replaced by the mouse position.
The additional video shows examples of other complex behaviors obtained for different parameter settings.
Figure \ref{fig:catmouse} (top-right) illustrates this behavior.

The types of solution we observe are different for other parameter values.
For example, for $M=2$ or a small time horizon, e.g. $H=1$, the dynamical state in which the cats rotate around the mouse is not
stable, and the mouse escapes. This is displayed in Figure \ref{fig:catmouse} (bottom panels) and better illustrated in the video provided as supplementary material. 
We emphasize that these different behaviors are observed for large uncertainty in the form of sensor noise and wind.

\section{Conclusions}
\label{sec:conc}
This paper presents a centralized, real-time stochastic optimal control algorithm for coordinating the actions of multiple autonomous vehicles in order to minimize a global cost function.
The high-level control task is expressed as a Path-Integral control problem that can be solved using efficient sampling methods and real-time control is possible via the use of re-planning and model predictive control. To the best of our knowledge, this is the first real-time implementation of Path-Integral control on an actual multi-agent system.

We have shown in a simple scenario (Drunken Quadrotor) that the proposed methodology is more convenient than other approaches such as deterministic control or iLQG for planning trajectories.
In more complex scenarios such as the Holding Pattern and the Cat and Mouse, the proposed methodology is also preferable and allows for real-time control. We observe multiple and complex group behavior emerging from the specified cost function. Our experimental framework \sname has been a key development that permitted a smooth transition from the theory to the real quadrotor platforms, with literally no modification of the underlying control code. This gives evidence that the model mismatch caused by the use of a control hierarchy is not critical in normal outdoor conditions.
Our current research is addressing the following aspects:

{\bf Large scale parallel sampling}$-$ the presented method can be easily parallelized, for instance, using graphics processing units, as in \citet{grady}.
Although the tasks considered in this work did not required more than $10^4$ samples, we expect that this improvement will significantly increase the number of application domains and system size.

{\bf Distributed control}$-$ we are exploring different distributed formulations that take better profit of the factorized representation of the state cost. Note that the costs functions considered in this work only require pairwise couplings of the agents (to prevent collisions). However, full observability of the joint space is still required, which is not available in a fully distributed approach.


\small
\bibliographystyle{aaai} 
\bibliography{references}









\end{document}